\documentclass[prl,twocolumn,showpacs]{revtex4-1}
\usepackage{amsmath,amscd,amsfonts,amssymb,color}
\usepackage{graphicx,amsfonts,epsf}
\begin{document}
\title{Experimentally freezing quantum discord in
a dissipative environment using dynamical decoupling}
\author{Harpreet Singh}
\email{harpreetsingh@iisermohali.ac.in}
\affiliation{Department of Physical Sciences, Indian
Institute of Science Education \& 
Research (IISER) Mohali, Sector 81 SAS Nagar, 
Manauli PO 140306 Punjab India.}
\author{Arvind}
\email{arvind@iisermohali.ac.in}
\affiliation{Department of Physical Sciences, Indian
Institute of Science Education \& 
Research (IISER) Mohali, Sector 81 SAS Nagar, 
Manauli PO 140306 Punjab India.}
\author{Kavita Dorai}
\email{kavita@iisermohali.ac.in}
\affiliation{Department of Physical Sciences, Indian
Institute of Science Education \& 
Research (IISER) Mohali, Sector 81 SAS Nagar, 
Manauli PO 140306 Punjab India.}
\begin{abstract}
The discovery of the intriguing phenomenon that certain
kinds of quantum correlations remain impervious to noise up
to a specific point in time and then suddenly decay, has
generated immense recent interest.  We exploit dynamical
decoupling sequences to prolong the persistence of
time-invariant quantum discord in a system of two NMR qubits
decohering in independent dephasing environments.  We
experimentally prepare two-qubit Bell-diagonal quantum
states that interact with  individual dephasing channels and
demonstrate the effect of dynamical decoupling on the
preservation of both quantum and classical correlations.  We
are able to freeze quantum discord over long time scales in
the presence of noise, using dynamical decoupling.  We use
robust state-independent dynamical decoupling schemes for
state preservation and demonstrate that these schemes are
able to successfully preserve quantum discord.
\end{abstract}
\maketitle
\noindent{\em Introduction:-} The quantification of quantum
correlations, distinction from their classical counterparts,
and their behavior under decoherence, is of paramount
importance to quantum information processing~\cite{nc-book}.
Several measures of nonclassical correlations have been
developed~\cite{lanyon-prl-08} and their signatures
experimentally measured on an NMR
setup~\cite{soarespinto-pra-10,silva-prl-13}.  Quantum
discord is a measure of nonclassical correlations that are
not accounted for by quantum
entanglement~\cite{olliver-prl-01}.  While the intimate
connection of quantum entanglement with quantum nonlocality
is well understood and entanglement has long been considered
a source of quantum computational
speedup~\cite{horodecki-rmp-09}, the importance of quantum
discord, its intrinsic quantumness and its potential use in
quantum information processing protocols, is being explored
in a number of contemporary studies~\cite{modi-rmp-12}.

A surprising recent finding that for certain quantum states
up to some time $\bar{t}$, quantum correlations are not
destroyed by decoherence whereas classical correlations
decay, and after this time $\bar{t}$ the situation is
reversed and the  quantum correlations begin to decay, has
generated widespread
interest~\cite{maziero-pra-09,mazzola-prl-10}.  This
inherent immunity of such quantum correlations to environmental noise
throws up new possibilities for the characterization of
quantum behavior and its exploitation for quantum
information processing.  The peculiar ``frozen'' behavior of quantum
discord in the presence of noise was experimentally
investigated using photonic qubits~\cite{xu-nc-10} and  NMR
qubits~\cite{auccaise-prl-11,paula-prl-13}.  The class of
initial quantum states that exhibit this sudden transition in
their decay rates was theoretically studied under the action
of standard noise channels, and it was inferred that the type of
states that display such behavior depends on the
nature of the decohering channel being
considered~\cite{fanchini-pra-10}.  Dynamical decoupling
methods have been proposed to protect quantum discord from
environment-induced errors~\cite{fanchini-pra-12} and  a recent
work showed that interestingly, dynamical decoupling schemes
can also influence the timescale over which time-invariant
quantum discord remains oblivious to
decoherence~\cite{addis-pra-15}.

We report the remarkable preservation of
time-invariant quantum discord upon applying
time-symmetric dynamical decoupling (DD) schemes
of the band-bang variety, on a two-qubit NMR
quantum information processor.  For DD schemes of the symmetric XY4 and
XY8 variety, quantum discord persists up to a time
$\bar{t} \approx 0.1$s and for  the symmetric
XY16 and KDD$_{xy}$ kind of DD schemes, the 
discord persists up to a time  $\bar{t} \approx
0.2$s which is double and four times respectively,
of the persistence time of $0.05$s when no
preservation schemes are applied.  This is the
first experimental demonstration of the potential
that  DD schemes hold for prolonging the lifetime of
time-invariant discord, as predicted in recent
theoretical work~\cite{addis-pra-15}.  We note
here, that while the theoretical results show that
the DD schemes can prolong the persistence of
time-invariant quantum discord for particular
kinds of environmental noise, in  our experiments,
different types of decohering channels could
contribute to spin decoherence.  Hence our results
reflect the remarkable success of DD schemes in
preserving quantum correlations, regardless of the
nature of the decohering channels that act on the
qubits. 
\begin{figure}[t]
\includegraphics[angle=0,scale=1.0]{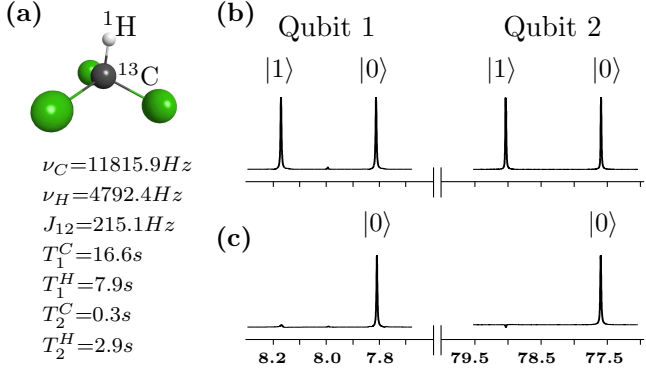}
\caption{(Color online) (a) 
Chloroform-${}^{13}$C molecular structure with 
${}^{1}H$ and ${}^{13}$C labeling the first and
second qubits, respectively.
Tabulated system parameters 
are:~chemical shifts $\nu_i$, the scalar coupling 
interaction strength $J_{12}$ (in
Hz) and relaxation times $T_{1}$ and $T_{2}$ (in seconds).
NMR spectrum of (b) the thermal equilibrium state after a $\pi/2$
readout pulse  and (c) the pseudopure $\vert 0 0 \rangle$
state.
}
\label{molecule}
\end{figure}
\noindent{\em Experimentally creating time-invariant
discord:-} We create and preserve time-invariant discord in a two-qubit
NMR system  of chloroform-$^{13}C$,
with the ${}^{1}$H and ${}^{13}$C nuclear spins encoding the
two qubits (Fig.~\ref{molecule}).  The ensemble of nuclear
spins is placed in a longitudinal strong static magnetic
field ($B_0 \approx 14.1$T) oriented along the $z$
direction.  The ${}^{1}$H and ${}^{13}$C nuclear spins
precess around $B_0$ at Larmor frequencies of $\approx 600$
MHz and $\approx 150$ MHz, respectively. The evolution of
spin magnetization is controlled by applying rf-field pulses
in the $x$ and $y$ directions.
The internal Hamiltonian of the system  in the rotating
frame is given by
\begin{equation}
H = \sum_{i=1}^{2} (\nu_i-\nu_{rf}^i) I_{iz} +  \sum_{i <
j, i=1}^{2} J_{ij} I_{iz}
I_{jz}
\end{equation}
where $\nu_i$ is the Larmor frequency of the $i$th spin,
$\nu_{rf}^i$  is the rotating frame frequency
($\nu_{rf}^i=\nu_i$ for on-resonance), and 
$J_{ij}$ is the spin-spin coupling constant. 
The two-qubit system
was initialized into the pseudopure state $\vert 00 \rangle$
by the spatial averaging technique~\cite{cory-physicad}.
Density matrices were reconstructed from
experimental data by using a reduced set of
quantum state tomography (QST)
operations~\cite{leskowitz-pra-04} combined with
the maximum likelihood method~\cite{singh-pla-16}
to avoid any negative eigen values.
The fidelity $F$ of all the experimental
density matrices reconstructed
in this work was computed
using the Uhlmann-Jozsa
measure~\cite{jozsa-fidelity}.
The experimentally created
pseudopure state $\vert 00 \rangle$ was tomographed with a
fidelity of $0.99$,  and the NMR signal of this state was
used as a reference for computation of state fidelity in all
subsequent time-invariant discord experiments.

A class of Bell-diagonal (BD) states with maximally mixed margins
are defined in terms of Pauli operators $\sigma_i$ as
\begin{equation}
\label{eqbd}
\rho_{\rm BD}=\frac{1}{4}\left(\,\mathbb{I} \right. +
{\sum_{i=1}^3} \left. c_i\ \sigma_i
\otimes\sigma_i \right)
\end{equation}
where $0 \le \vert c_i \vert \le 1$ determine the
state completely and can be computed 
as $c_i = \langle \sigma_i \otimes \sigma_i
\rangle$.

\begin{figure}[hbtp] 
\centering
\includegraphics[angle=0,scale=1.0]{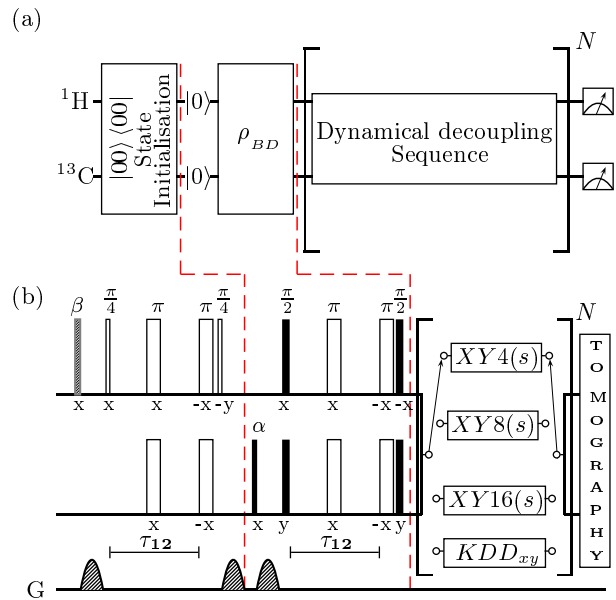}
\caption{(Color online)
(a) Quantum circuit for the initial pseudopure
state preparation, followed by the block for 
BD state preparation. The next block depicts
the DD scheme used to
preserve quantum discord. The
entire DD sequence is repeated $N$ times before
measurement.
(b) NMR pulse sequence corresponding to the
quantum circuit. The rf pulse flip angles are set
to $\alpha=46^{\circ}$ and $\beta= 59.81^{\circ}$,
while all other pulses are labeled with their
respective angles and phases.
}
\label{ckt}
\end{figure}

We aim to prepare an initial BD state
with the  parameters $c_1(0) = 1$, $c_2(0) = 0.7$,
$c_3(0) = -0.7$.  The NMR pulse
sequence for the preparation of this state 
from the  $\vert 00 \rangle$ pseudopure state
is given in Fig.~\ref{ckt}(b).  
Preparing the BD state involves 
manipulation of NMR multiple-quantum coherences by applying
rotations in the zero-quantum and double-quantum 
spin magnetization subspaces.
Since the molecule
is a heteronuclear spin system,
high-power, short-duration rf pulses were used for
gate implementation (with rf pulses of flip
angles $\alpha=46^{\circ}$ and $\beta=
59.81^{\circ}$ in Fig.~\ref{ckt} having pulse
lengths of $6.85 \mu$s and $5.02 \mu$s,
respectively).  The experimentally achieved
$\rho^{\rm E}_{\rm BD}$ (reconstructed using the
maximum likelihood method~\cite{singh-pla-16}) had
parameters $c_1(0) = 1.0$, $c_2(0) = 0.708$ and
$c_3(0) = -0.708$ and a computed fidelity of
$0.99$.  The experimentally reconstructed density
matrix (using quantum state tomography and maximum
likelihood) was found to be:

\begin{widetext}
\begin{equation}
\rho^{\rm E}_{\rm BD} =
\left(\begin{array}{cccc}
0.0724 + 0.0000i&-0.0035 + 0.0026i & -0.0035 + 0.0026i &
0.0730 + 0.0003i \\
-0.0035 - 0.0026i &  0.4268 + 0.0000i &  0.4270 + 0.0006i &
-0.0028 -
0.0020i \\
-0.0035 - 0.0026i &  0.4270 - 0.0006i &  0.4273 + 0.0000i &
-0.0028 -
0.0020i \\
   0.0730 - 0.0003i & -0.0028 + 0.0020i & -0.0028 + 0.0020i
&  0.0735 +
0.0000i
\end{array}
\right)
\end{equation}
\end{widetext}

Correlations function for the BD states
can be computed readily to give us
the classical correlations (${\cal C}[\rho(t)]$),
the quantum discord (${\cal D}[\rho(t)]$), and
total correlations (${\cal
I}[\rho(t)]$~\cite{mazzola-prl-10}: 
\begin{eqnarray}
{\cal C}[\rho(t)]&=&\sum_{j=1}^{2}
\frac{1+(-1)^{j}\chi(t)}{2}\log_{2}[1+(-1)^{j}\chi(t)] 
\nonumber \\
{\cal I}[\rho(t)]&=&
\sum_{j=1}^{2}\frac{1+(-1)^{j}
c_1(t)}{2}\log_{2}[1+(-1)^{j}c_1(t)] \nonumber \\
&&+
\sum_{j=1}^{2}\frac{1+(-1)^{j}
c_3}{2}\log_{2}[1+(-1)^{j}c_3]
\nonumber \\
{\cal D}(\rho)&\equiv&{\cal I}(\rho)-{\cal C}(\rho) 
\label{relations}
\end{eqnarray}
where $\chi(t)=\max\{|c_{1}(t)|,|c_{2}(t)|,|c_{3}(t)|\}$.

For the class of BD states with coefficients
$c_1=\pm 1, c_2=\mp c_3, \vert c_3 \vert < 1$,
and that decohere under a phase damping channel with
damping rate $\gamma$, the 
coefficients  evolve in time as
$c_1(t)=c_1(0)\exp{[-2\gamma t]}, c_2(t) = c_2(0)
\exp{[-2\gamma t]}, c_3(0) \equiv c_3$, and   
the quantum discord does not decay up to some
finite time $\bar{t}$~\cite{mazzola-prl-10}.  
We allowed the experimentally prepared state 
$\rho^{\rm E}_{\rm BD}$, 
to evolve freely in time and determined the
parameters $c_i$ at each time point.  We used these
experimentally determined coefficients
to compute the classical correlations (${\cal
C}[\rho(t)]$), the quantum discord (${\cal
D}[\rho(t)]$), and total correlations (${\cal
I}[\rho(t)]$) at each time point.  

\begin{figure} 
\centering
\includegraphics[angle=0,scale=1.0]{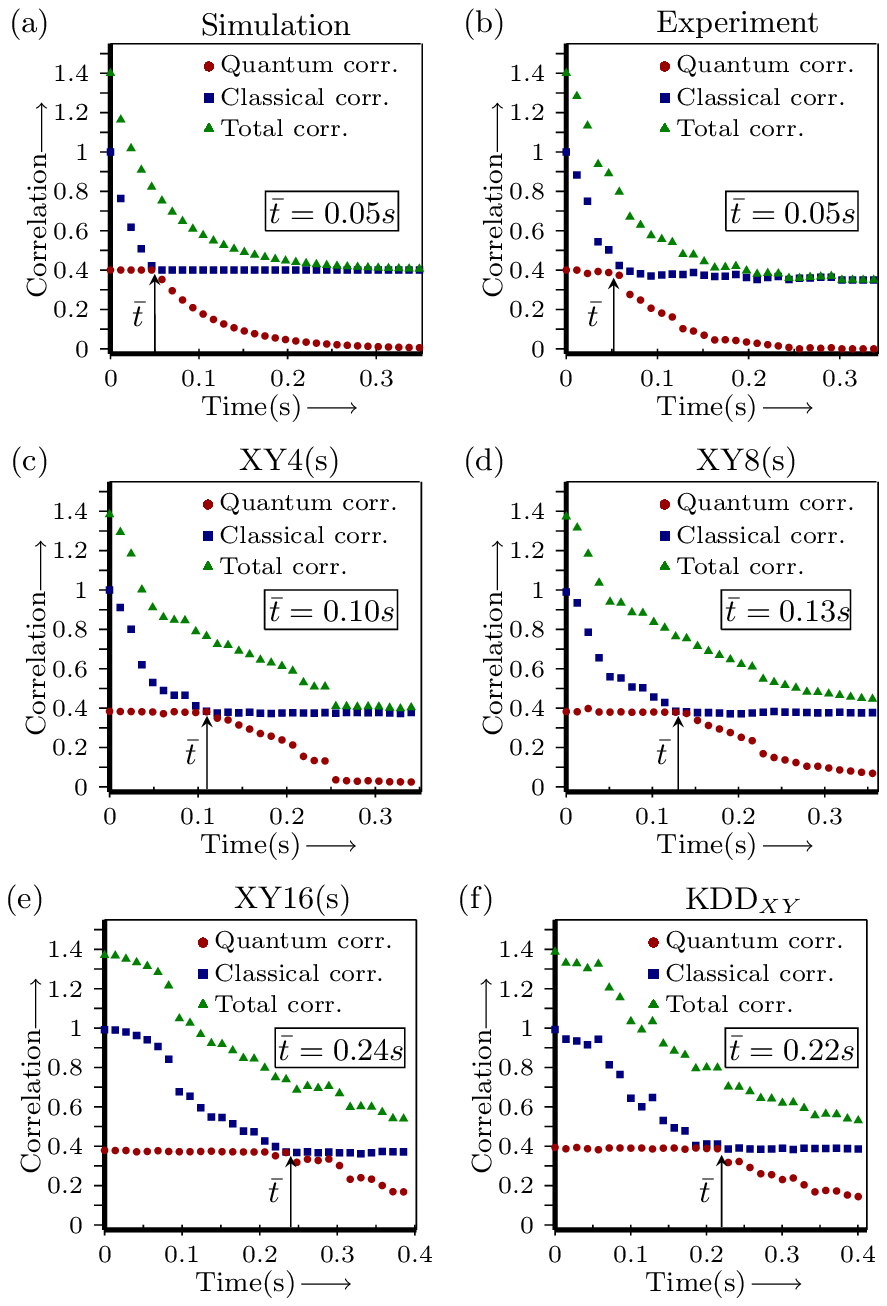}
\caption{(Color online) Time evolution of 
total correlations (green triangles), 
classical correlations (blue squares) 
and quantum discord (red circles) of 
the BD state:  (a) Simulation, (b)
Experimental plot without applying any preservation, and
(c)-(f) Experimental plots using preserving DD
sequences 
XY4(s), XY8(s), XY16(s) and KDD$_{xy}$,
respectively.
The displayed $\bar{t}$ values 
clearly
show 
the remarkable preservation of quantum discord 
under DD.
}
\label{tidplot_2}
\end{figure}

\begin{figure}
\centering
\includegraphics[angle=0,scale=1.0]{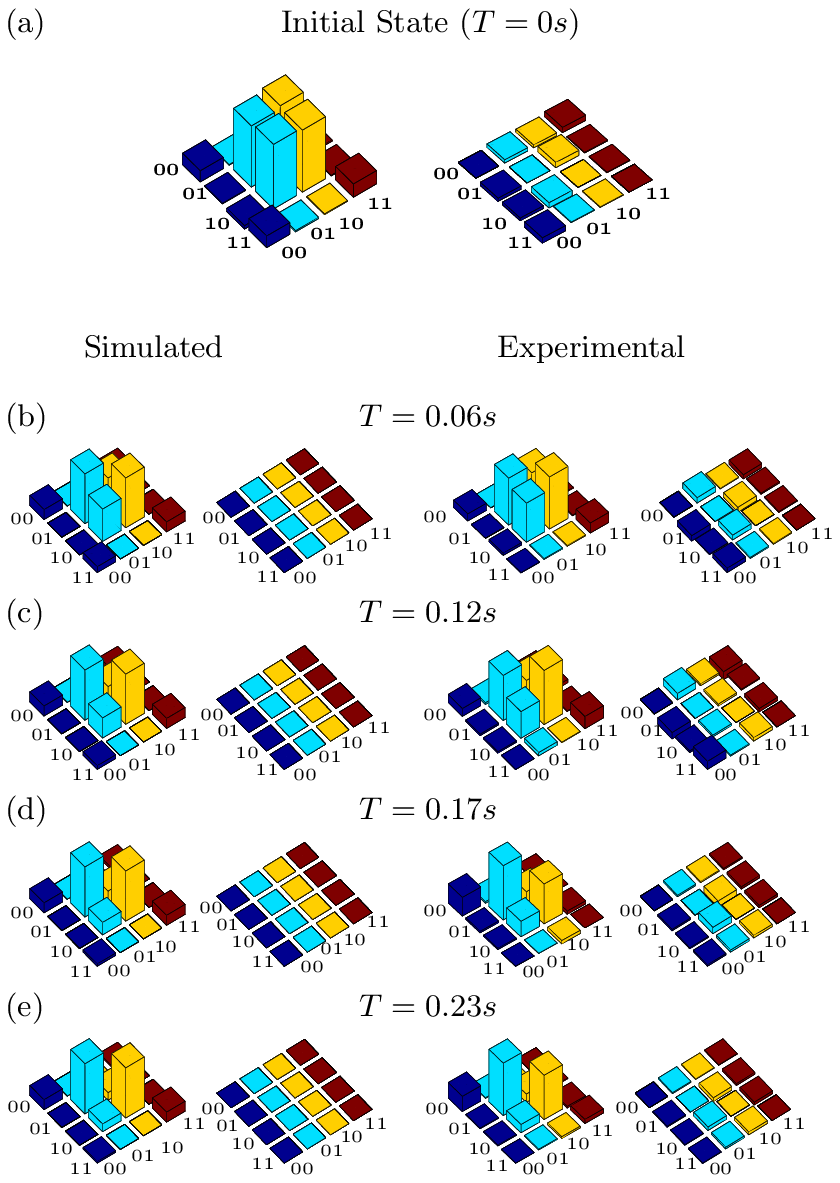}
\caption{
Real (left) and imaginary (right) parts of the experimental
tomographs of the (a) Bell Diagonal (BD) state, with a computed
fidelity of 0.99.  (b)-(e) depict the state at $T = 0.06,
0.12, 0.17, 0.23 $s, with the tomographs on the left and the
right representing the simulated and experimental state, respectively. 
The rows and columns are labeled in the computational basis ordered from
$\vert 00 \rangle$ to $\vert 11 \rangle$.
}
\label{nodd1}
\end{figure}

\begin{figure}
\centering
\includegraphics[angle=0,scale=1.0]{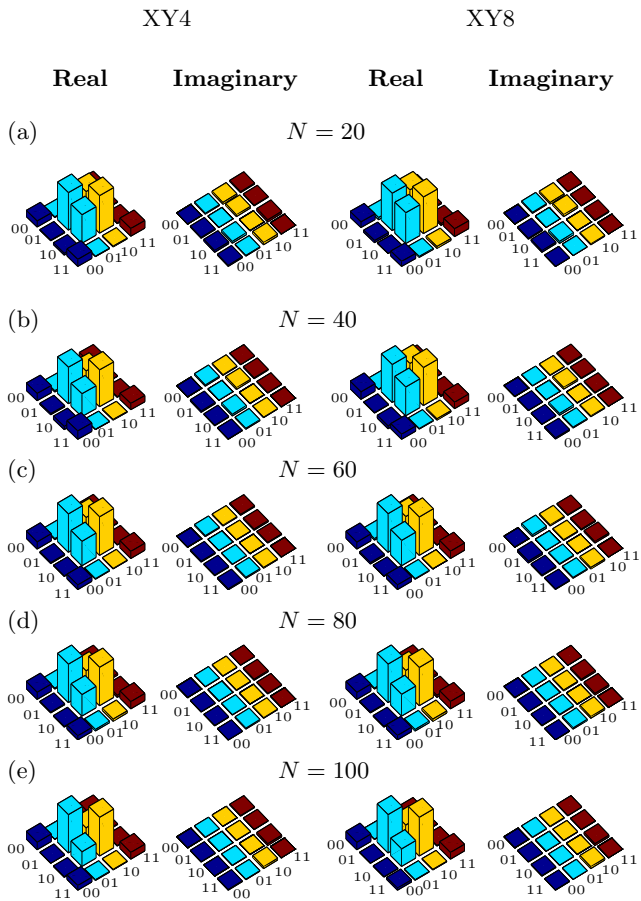}
\caption{
Real (left) and imaginary (right) parts of the experimental
tomographs of the (a)-(e) depict the BD state at $N = 20,
40, 60, 80,100 $, with the tomographs on the left and the
right representing the BD state after applying the XY4 and 
XY8 scheme, respectively.  The rows and
columns are labeled in the computational basis ordered from
$\vert 00 \rangle$ to $\vert 11 \rangle$.
}
\label{nodd2}
\end{figure}

\begin{figure}
\centering
\includegraphics[angle=0,scale=1.0]{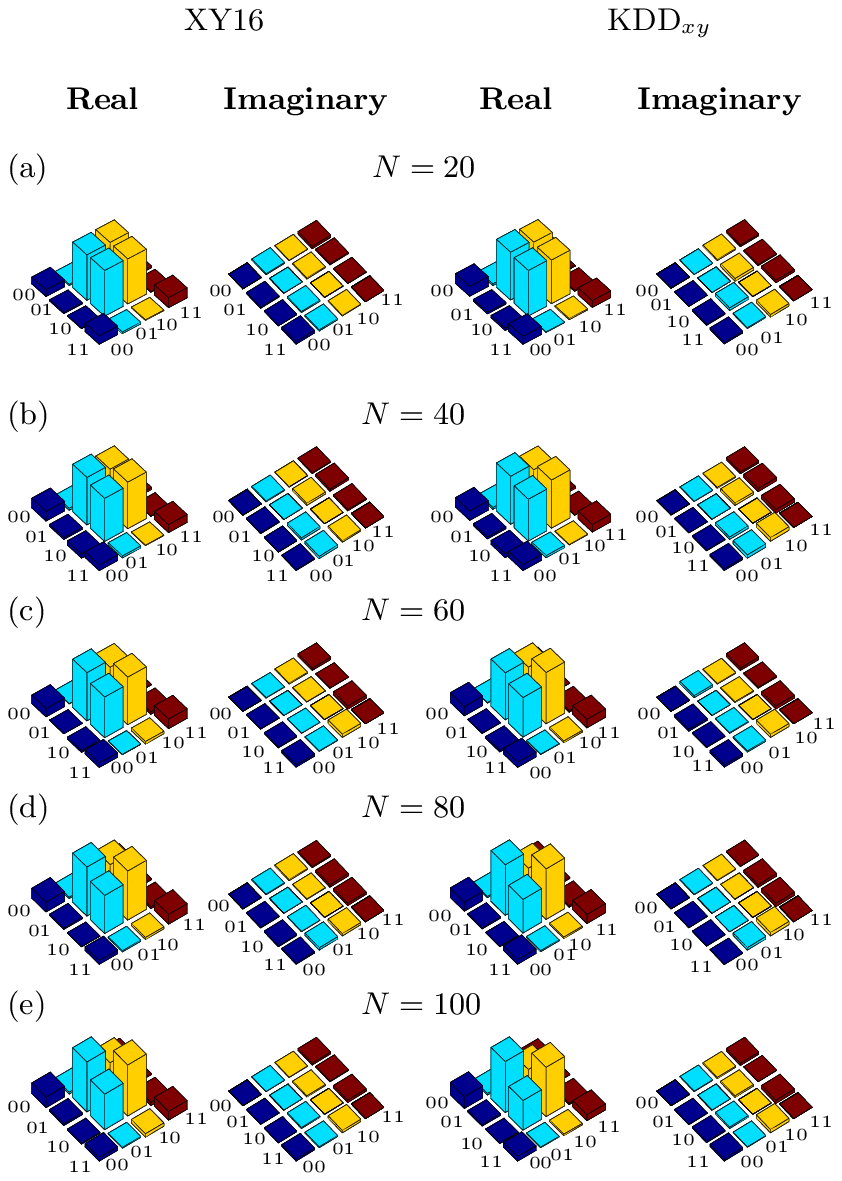}
\caption{Real (left) and imaginary (right) parts of the experimental
tomographs  in (a)-(e) depict the Bell Diagonal (BD) state at $N = 20,
40, 60, 80,100 $, with the tomographs on the left and the
right representing the BD state after applying the XY16 and 
KDD$_{xy}$ preserving DD schemes, respectively.  The rows and
columns are labeled in the computational basis ordered from
$\vert 00 \rangle$ to $\vert 11 \rangle$.
}
\label{nodd3}
\end{figure}

The experimental results are consistent with 
a model where the phase damping rate for the two-qubit 
system (carbon $C$,
proton $H$)
is modeled as $\gamma =
\frac{T_2^{*H}+T_2^{*C}}{2T_2^{*H} T_2^{*C}}$,
where the parameter $T_2^{*}$ accounts for rf
inhomogeneity in the static magnetic field as well
as for spin-spin relaxation.  Initially ($t=0$),  the
computed correlations in the $\rho^{\rm E}_{\rm
BD}$ state (taken as an average of five density
matrices initially prepared in the same state)
turned out to be ${\cal C}[\rho(0)]= 1.117 \pm
0.008 $, ${\cal D}[\rho(0)]= 0.379 \pm 0.007$ and
${\cal I}[\rho(0)]= 1.496 \pm 0.012$.  The
simulated and experimental plots of the dynamics
of the quantum discord, classical correlations and
total correlations are shown in
Fig.~\ref{tidplot_2} (a) and (b) respectively, and
show a distinct transition from the classical to
the quantum decoherence regimes at $t = \bar{t}$.
The experimentally determined phase damping rates
turn out to be $T_2^{*H}=0.41$ s and $T_2^{*C}=
0.19$ s.  The transition time up to which quantum
discord remains constant is
$\bar{t}=\frac{1}{2\gamma} \ln
\left|\frac{c_{1}(0)}{c_{3}(0)}\right|$, and is
experimentally determined to be $\bar{t} = 0.052$
s.  The experimentally tomographed density matrix,
reconstructed at different time points, shows that
state evolution remains confined to the subspace
of BD states, as is evident from the
experimentally reconstructed density matrices of the
state at different time points (with and without
DD preservation) displayed in Fig.\ref{nodd1} - Fig.\ref{nodd3},
respectively.

\noindent{\em Protecting time-invariant
discord:-}
Dynamical decoupling (DD) schemes, consisting of
repeated sets of $\pi$ pulses with tailored
inter-pulse delays and phases, have played
an important role in dealing with the debilitating effects of
decoherence~\cite{uhrig-prl-09}. Several NMR QIP
experiments have successfully used DD-type schemes to
preserve quantum states~\cite{roy-pra-11,singh-pra-14}.
\begin{figure}[hbtp] 
\centering
\includegraphics[angle=0,scale=1.0]{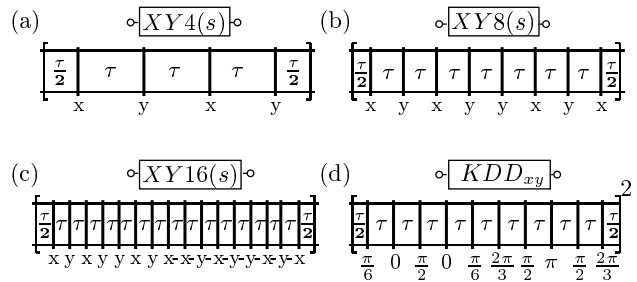}
\caption{
NMR pulse sequence corresponding to DD schemes
(a) XY4(s), (b) XY8(s), (c) XY16(s), and (d) KDD$_{xy}$. 
All the pulses are of flip angle $\pi$ and are labeled
with their respective phases. The pulses act on both 
qubits simultaneously.}
\label{ddpulses}
\end{figure}

The effect of time-reversal symmetry in dynamical
decoupling is extensively discussed
in~\cite{souza-prl-11}, wherein $\pi$ pulses along
both the $x$- and $y$- axes reduce errors due to
pulse imperfections; the XY4 sequence, for
instance, consists of four pulses with different
phases.  Pulse errors are almost perfectly
compensated if symmetric building blocks of $\pi$
pulses are used in a higher-order DD sequence, and
the XY8 and XY16 are symmetrized versions of the
XY4 sequence~\cite{ahmed-pra-13}.  A robust
sequence KDD$_{xy}$ designed to compensate for
flip-angle and resonance-offset errors, uses a
composite $\pi$ pulse~\cite{ahmed-pra-13}.  We used
time-symmetric building blocks in all the four
DD schemes implemented in this work:
$XY4(s)$, $XY8(s)$, $XY16(s)$ and KDD$_{xy}$,  to
preserve time-invariant discord
(Fig.~\ref{ddpulses}).  We applied $\pi$ pulses
simultaneously on both spins, with pulse lengths
of 15.1$\mu$s and 26.8 $\mu$s for the proton and
carbon spins,  respectively.  Each DD sequence was
applied a repeated number of times ($N$ being as
large as experimentally possible), for good
preservation.

Four $\pi$ pulses are required to implement the XY4 DD scheme,
$[\frac{\tau}{2}-\pi_{x} -\tau -\pi_{y}- 
\tau -\pi_{x}-\tau -\pi_{y}- \frac{\tau}{2}]_N$. 
We implemented the symmetrized XY4 DD
scheme for $\tau= 0.58$ ms and an
experimental 
time for
one run of $2.43$ ms.
The time for which quantum discord persists using
the XY4(s) scheme is
$\bar{t}=0.11$s, which is double the time
as compared to no-preservation (Fig.~\ref{tidplot_2}(c)).  
We implemented the
XY8 DD
scheme 
$[\frac{\tau}{2}-\pi_{x} -\tau -\pi_{y}- \tau -\pi_{x}- 
\tau -\pi_{y}--\tau -\pi_{y}- \tau -\pi_{x}- \tau -
\pi_{y}-\pi_{x}- \frac{\tau}{2}]_N$ ( 
providing a better
compensation for pulse errors),  
for $\tau$= 0.29 ms and an experimental time for
one run of $2.52$ ms. 
The time for which quantum discord
persists using the XY8(s) scheme
is $\bar{t}=0.13$s, nearly the same as
the XY4 scheme (Fig.~\ref{tidplot_2}(d)).  
The
XY16(s) DD scheme $[\frac{\tau}{2}-\pi_{x}-\tau
-\pi_{y}-\tau -\pi_{x}-\tau -\pi_{y}-\tau -\pi_{y}- \tau
-\pi_{x}-\tau -\pi_{y}-\tau -\pi_{x } -\tau -\pi_{-x}-\tau
-\pi_{-y}-\tau -\pi_{-x}-\tau - \pi_{-y}-\tau -\pi_{-y}-\tau
-\pi_{-x}-\tau -\pi_{-y}-\tau -\pi_{-x }-
\frac{\tau}{2}]_N$ provides even better compensation
than the XY8 sequence.  
We implemented the XY16(s)
scheme for $\tau$= 0.145 ms and an
experimental time for one run of $2.75$ ms.
The time for which quantum discord
persists for the XY16(s) scheme is
$\bar{t}=0.24$s, which is four times the
persistence time of the discord when no
preservation is applied (Fig.~\ref{tidplot_2}(e)).
The basic KDD sequence is given by KDD$_\phi$ =
$\tau/2-(\pi)_{\pi/6+\phi}-\tau-(\pi)_{\phi}-\tau-
(\pi)_{\pi/2+\phi}-\tau-(\pi)_{\phi}-\tau-(\pi)_{\pi/6+\phi}-\tau/2
$. To improve robustness, the sequence can be extended by
combining blocks of five pulses shifted in phase by $\pi/2$,
such as [KDD$_\phi$ - KDD$_{\phi+\pi/2}$]$^2$, where the
lower index denotes the overall phase of the block. 
We implemented the KDD$_{xy}$ sequence~\cite{ahmed-pra-13}:
$[\tau/2-(\pi)_{\pi/6}-\tau-(\pi)_{0}-\tau-(\pi)_{\pi/2}-
\tau-(\pi)_{0}-\tau-(\pi)_{\pi/6}-\tau-(\pi)_{2\pi/3}-\tau-
(\pi)_{\pi}-\tau-(\pi)_{\pi}-\tau-
(\pi)_{\pi/2}-\tau-(\pi)_{2\pi/3}-\tau/2]^2$, with 
$\tau$= 0.116 ms and an experimental time for one
run of 2.86 ms.
In this case too, 
the time for which quantum
discord persists $\bar{t}=0.22$s is quadrupled, as
compared to no-preservation
(Fig.~\ref{tidplot_2}(f)).  
The DD sequence was looped for five times between each
time point in Figs.~\ref{tidplot_2}(c)-(f), which typically 
means repeating a DD sequence 175-200 times during an
experiment covering all time points.
Details of the experimentally
tomographed density matrices reconstructed at
different time points, after preservation by
different DD schemes, are discussed
in~\cite{harpreet-supple}.
Our results show that time-invariant quantum discord, which
remains unaffected under certain decoherence regimes, can be
preserved for very long times using dynamical decoupling
schemes.  Our experiments have important implications in
situations where persistent quantum correlations have to be
maintained in order to carry out quantum information
processing tasks.

\begin{acknowledgments}
All experiments were performed on a Bruker
Avance-III 600 MHz FT-NMR spectrometer at the NMR
Research Facility at IISER Mohali.  KD
acknowledges funding from DST India under Grant
No. EMR/2015/000556.  Arvind acknowledges funding
from DST India under Grant No. EMR/2014/000297.
HS acknowledges CSIR India for financial support.
\end{acknowledgments} 
\end{document}